\documentclass{article}

\usepackage{arxiv}

\usepackage[utf8]{inputenc} % allow utf-8 input
\usepackage[T1]{fontenc}    % use 8-bit T1 fonts
\usepackage{hyperref}       % hyperlinks
\usepackage{url}            % simple URL typesetting
\usepackage{booktabs}       % professional-quality tables
\usepackage{amsfonts}       % blackboard math symbols
\usepackage{nicefrac}       % compact symbols for 1/2, etc.
\usepackage{microtype}      % microtypography
\usepackage{lipsum}
\usepackage{graphicx}
\usepackage{framed,multirow}
\usepackage{tabularx}
\graphicspath{ {./images/} }
\usepackage{amssymb}
\usepackage{amsmath}

\title{DM-SegNet: Dual-Mamba Architecture for 3D Medical Image Segmentation with Global Context Modeling}

\author{
 Hangyu Ji \\
  School of Computer Science and Engineering\\
  Central South University\\
  ChangSha, 410083 \\
  \texttt{234712170@csu.edu.cn} \\
}

\begin{document}
\maketitle
\begin{abstract}
%% Text of abstract
Accurate 3D medical image segmentation demands architectures capable of reconciling global context modeling with spatial topology preservation. While State Space Models (SSMs) like Mamba show potential for sequence modeling, existing medical SSMs suffer from encoder-decoder incompatibility: the encoder's 1D sequence flattening compromises spatial structures, while conventional decoders fail to leverage Mamba's state propagation. We present DM-SegNet, a Dual-Mamba architecture integrating directional state transitions with anatomy-aware hierarchical decoding. The core innovations include a quadri-directional spatial Mamba module employing four-directional 3D scanning to maintain anatomical spatial coherence, a gated spatial convolution layer that enhances spatially sensitive feature representation prior to state modeling, and a Mamba-driven decoding framework enabling bidirectional state synchronization across scales. Extensive evaluation on two clinically significant benchmarks demonstrates the efficacy of DM-SegNet: achieving state-of-the-art Dice Similarity Coefficient (DSC) of 85.44\% on the Synapse dataset for abdominal organ segmentation and 90.22\% on the BraTS2023 dataset for brain tumor segmentation.
\end{abstract}

% keywords can be removed
%\keywords{First keyword \and Second keyword \and More}
\keywords{Mamba \and 3D Medical Image Segmentation \and State Space Models \and Volumetric Analysis}

\section{Introduction}
\label{sec1}
%% Labels are used to cross-reference an item using \ref command.
Accurate 3D medical image segmentation is foundational to computer-aided diagnosis, yet processing clinical-scale volumes remains computationally prohibitive for existing architectures\cite{niyas_MedicalImageSegmentation_2022,litjens_SurveyDeepLearning_2017,shorten_SurveyImageData_2019}. While convolutional networks like 3D U-Net \cite{cicek_3DUNetLearning_2016} establish local feature hierarchies through cascaded downsampling, their constrained receptive fields fundamentally limit global dependency modeling. Though Vision Transformers (UNETR \cite{hatamizadeh_UNETRTransformers3D_2021}, SwinUNETR \cite{cao_SwinUnetUnetlikePure_2021}) address this via self-attention mechanisms, their quadratic complexity makes them impractical for high-resolution 3D segmentation\cite{vaswani_AttentionAllYou_2023}.

State Space Models (SSMs), particularly Mamba \cite{gu_EfficientlyModelingLong_2022,gu_MambaLinearTimeSequence_2024}, emerge as promising alternatives by enabling linear-time global context modeling. Early medical adaptations like U-Mamba \cite{ma_UMambaEnhancingLongrange_2024} demonstrate memory reduction over Transformers through axial sequence flattening, but at the cost of disrupting 3D spatial topology—critical anatomical continuity. More critically, existing Mamba variants inherit a structural dichotomy: encoders leverage state transitions for sequence modeling, while decoders regress to conventional transposed convolutions that discard learned state dependencies. This encoder-decoder incompatibility causes progressive context dissipation during upsampling, particularly impairing low-contrast lesion delineation.

We propose DM-SegNet, a dual-Mamba architecture that unifies selective state propagation with anatomically coherent feature decoding. Our key contributions address three core limitations:
\begin{enumerate}
\item Direction-aware state scanning: A quadri-directional state Mamba (QSM) module preserves 3D spatial relationships through four-directional (axial/sagittal/coronal + reverse) parallel scanning, eliminating topology distortion from 1D sequence flattening.
\item Spatially sensitive feature enhancement: Gated spatial convolution (GSC) layers prior to Mamba blocks adaptively enhance spatially-aware feature propagation.
\item Decoding enhancement: The multi-scale fusion Mamba decoder introduces cross-scale dependency propagation through gated Mamba blocks and maintains cross-scale context consistency by bidirectionally aligning the encoder state with the decoder feature mapping.

\end{enumerate}

On Synapse and BraTS2023 benchmarks, DM-SegNet achieves state-of-the-art Dice Similarity Coefficient (DSC) of 85.44\% on the Synapse dataset for abdominal organ segmentation and 90.22\% on the BraTS2023 dataset for brain tumor segmentation.

\section{Related Work}
\paragraph{3D Convolutional Networks}
 Pioneering architectures like 3D U-Net \cite{cicek_3DUNetLearning_2016} established encoder-decoder frameworks for volumetric segmentation through symmetric downsampling and skip connections\cite{taha_MetricsEvaluating3D_2015}. While effective in local feature aggregation, their limited receptive fields struggle to capture long-range dependencies critical for infiltrative pathologies. However, efforts to expand the convolution kernel size to achieve a large receptive field, like those in 3D UX-Net \cite{lee_3DUXNetLarge_2023}, have led to high memory costs.
\paragraph{Vision Transformers}
The introduction of self-attention mechanisms via UNETR \cite{hatamizadeh_UNETRTransformers3D_2021} enabled global context modeling, achieving breakthroughs in volume brain tumor and spleen segmentation. SwinUNETR \cite{cao_SwinUnetUnetlikePure_2021}, though its hierarchical architecture mitigates the traditional Transformer's position encoding flaw, may still suffer from segmentation accuracy drop in fine - grained localization tasks due to position information loss.
\paragraph{State Space Models}
Mamba-based architectures \cite{gu_MambaLinearTimeSequence_2024} has become the latest choice for both accuracy and efficiency, with U-Mamba \cite{ma_UMambaEnhancingLongrange_2024} significantly reduces memory usage compared to the Transformer through axial sequence modeling. However, flattening 3D volumes into 1D sequences disrupts anatomical topology, degrading performance on continuous structures.
\paragraph{Decoder Architectures}
The core contradiction in current decoder designs lies in the semantic granularity mismatch between the high-level semantic features output by the encoder and the pixel-level localization information required by the decoder. Although frameworks like UNETR \cite{hatamizadeh_UNETRTransformers3D_2021}  pass multi-scale features through skip connections, their direct concatenation strategy neglects the representational gap between the abstract semantics of the encoder and the localization needs of the decoder. Recent SSM-based methods such as SegMamba \cite{xing_SegMambaLongrangeSequential_2024}, while achieving efficient global modeling in the encoder, still rely on traditional transposed convolutions in the decoder.

\begin{figure}[t]
     \centering
     \includegraphics[width=1\columnwidth]{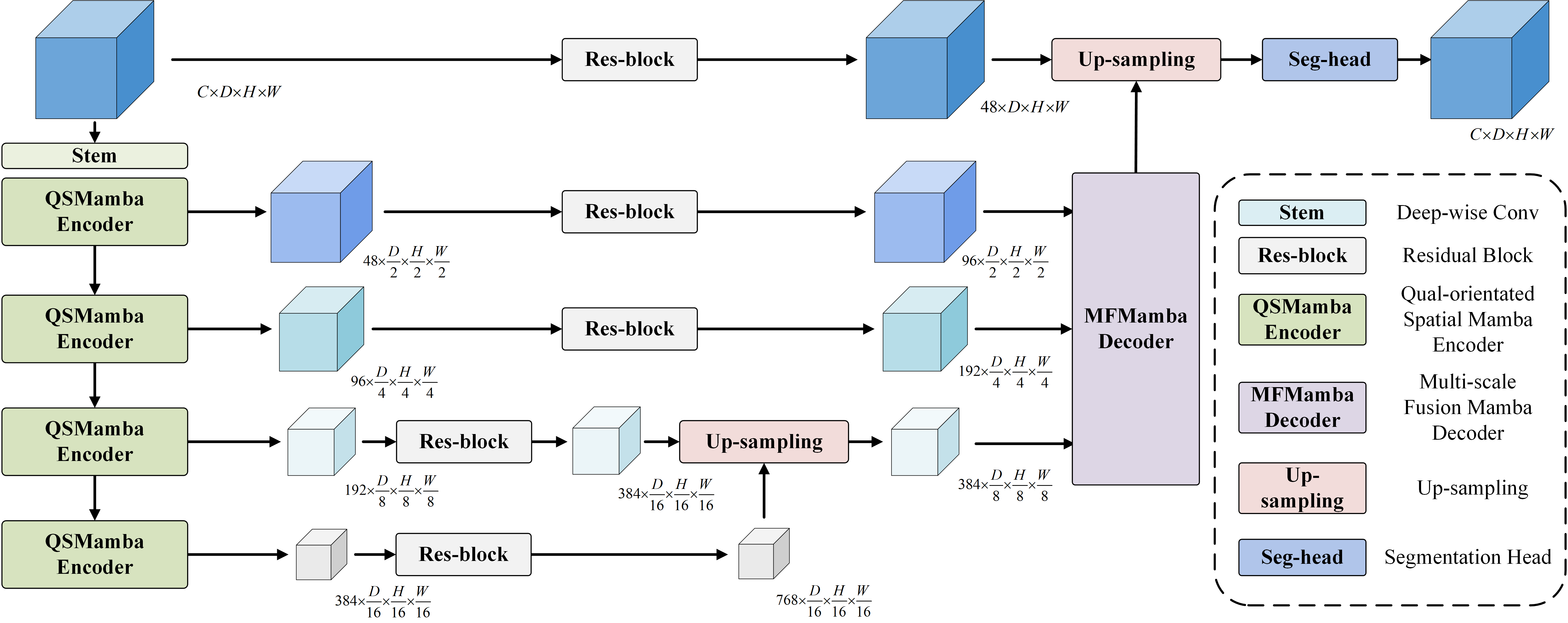}
     \caption{Overview of DM-SegNet architecture. The encoder architecture comprises a stem convolutional block followed by four hierarchical QSMamba encoder modules. These modules gradually extract multi-scale feature representations through the QSMamba modules and downsampling. The encoded features are then passed through the residual path to the MFMamba decoder, thereby achieving the collaborative fusion of local structural details and global context dependencies. The decoder ultimately generates pixel-wise segmentation predictions through iterative feature recalibration and spatial resolution recovery operations.}\label{fig1}
\end{figure}

\section{Method}
\subsection{Overall Architecture}
As shown in Figure \ref{fig1}, our proposed DM-SegNet consists of a quadri-directional spatial Mamba feature encoder and a novel multi-scale fusion Mamba decoder. The quadri-directional spatial Mamba encoder effectively captures long-range dependencies in volumetric features at each scale, while the multi-scale fusion Mamba decoder enables multi-scale feature fusion and global feature modeling while maintaining high efficiency during both training and inference. Overall, the Mamba encoder injects powerful global context and local details into the feature pyramid, where feature maps undergo progressive downsampling and transformation to generate full-scale features at different pyramid levels. These features are then aggregated through upsampling modules with skip connections and processed by the Mamba decoder along with the full-scale features. The resulting features containing comprehensive global spatial information are combined with local detail features extracted from the original image, ultimately generating integrated multi-scale representations for dense prediction through global feature modeling.
\subsection{Quadri-directional Spatial Mamba Encoder}
\begin{figure}[t]
     \centering
     \includegraphics[width=1\columnwidth]{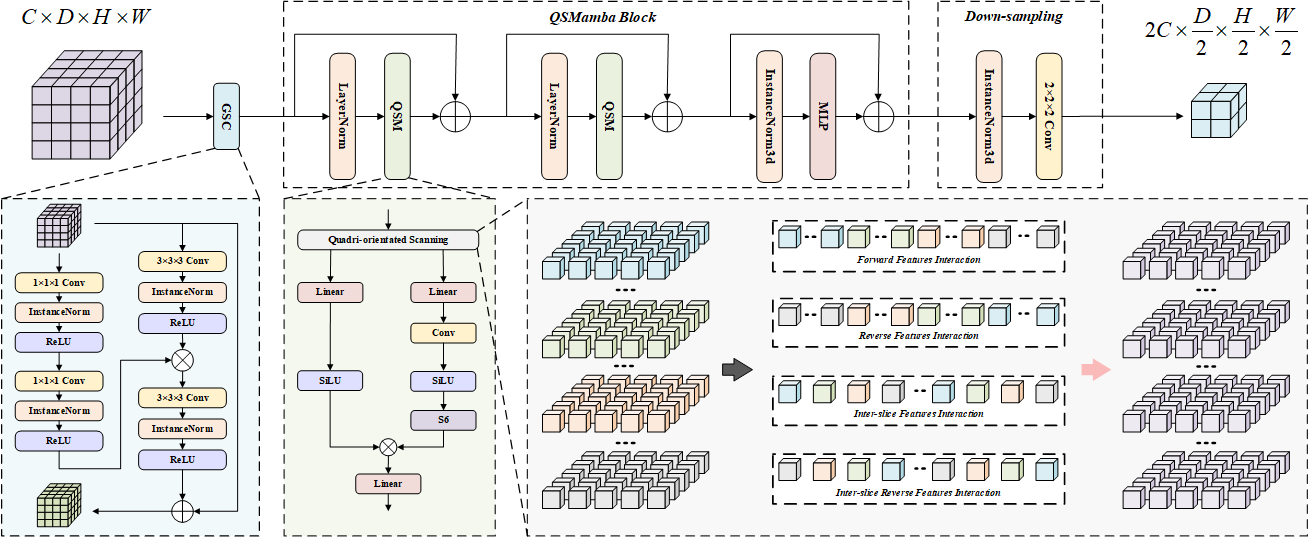}
     \caption{QSMamba Encoder structure diagram. It consists of a Gated Spatial Convolution module, multiple Quadri-directional Spatial Mamba modules and a Down-sampling module.}\label{fig2}
\end{figure}

\paragraph{Overview}
As shown in Figure \ref{fig2}, our Quadri-directional Spatial Mamba Encoder (QSMamba Encoder) consists of a Gated Spatial Convolution (GSC), multiple QSMamba blocks, and a Down-sampling module. A GSC is inserted before each QSMamba block to suppress noisy channels via a gating mechanism, compensates for spatial information lost during Mamba’s 1D sequence flattening, and enhances local context. A Down-sample module is appended after each QSMamba block to reduce memory and computational costs while providing multi-scale features for subsequent decoder stages.

\paragraph{Gated Spatial Convolution}
Conventional Mamba architectures flatten 3D feature tensors \( z \in \mathbb{R}^{C \times H \times W \times D} \) into 1D sequences for processing, leading to the loss of explicit spatial topology modeling capabilities. To address this, we propose the Gated Spatial Convolution module, which reconstructs spatial awareness through the synergy of convolutional inductive bias and dynamic gating.

As shown in Figure \ref{fig2}, the GSC module contains two main branches and a gating mechanism:

The first branch is the feature transformation branch, which first uses a $1 \times 1 \times 1$ convolution to perform channel-wise transformation on the input 3D features, adjusting inter-channel correlations without altering spatial dimensions. This is followed by a second $1 \times 1 \times 1$ convolution to further enhance nonlinear representation capabilities, allowing more comprehensive fusion and expression of features along the channel dimension.

The second branch employs two sequential $3 \times 3 \times 3$ convolution operations: the first $3 \times 3 \times 3$ convolution extracts local spatial features to expand the receptive field, while the second $3 \times 3 \times 3$ convolution further strengthens spatial feature extraction to deeply mine spatial information. Each convolution is followed by Instance Normalization (IN) and ReLU activation to ensure data stability and nonlinear feature extraction.

The outputs of the left and right branches are then combined by element-wise multiplication  to form gating signals. Finally, the input features are fused with the gated features via element-wise addition  to generate the final output features.
\begin{equation}
     \begin{aligned}
          \mathbf{G}(z)_1 &= \text{ReLU}(\text{IN}(z * W_1^g)) \\
          \mathbf{G}(z)_2 &= \text{ReLU}(\text{IN}(\mathbf{G}(z)_1 * W_1^g))
     \end{aligned}
     \end{equation}
\begin{equation}
     \mathbf{C}(z) = \text{ReLU}(\text{IN}(z * W_3^g))
\end{equation}
\begin{equation}
     \mathrm{GSC}(z) = z + \mathbf{C}(z) \odot \mathbf{G}(z)_2
\end{equation}
where $z$ denotes the input 3D features, $W_n^g$ denotes the convolution block, $\mathrm{GSC}(z)$ denotes the final output.

\paragraph{Quadri-directional Spatial Mamba}
The original Mamba architecture was designed for processing 1D sequential data (e.g., text or speech) and is inherently unsuitable for high-dimensional medical imaging. To effectively leverage Mamba's advantages in feature extraction while handling medical volumetric data, we adapt it through dimension reduction strategies. Given the complexity of medical images, such as the heterogeneity of brain tumor shapes and the size differences among abdominal organs, the traditional parallel scanning strategy performs poorly when dealing with these complex feature differences. Therefore, an improved scanning strategy is needed to better adapt to medical image processing.

Existing visual Mamba networks employ various scanning directions including parallel, snake, bidirectional, spatial, and diagonal patterns \cite{liu_VMambaVisualState_2024,yang_PlainMambaImprovingNonHierarchical_2024,zhu_VisionMambaEfficient_2024}. To balance accuracy and computational efficiency, we propose a quadri-directional scanning strategy that constructs a four-way feature interaction network comprising(Figure \ref{fig2}): Forward Scanning, Reverse Scanning, Inter-slice Scanning, Reverse Inter-slice Scanning. Each directional path performs global context modeling through 2D selective scanning, followed by channel-wise concatenation and gated fusion to generate spatially enhanced features. This design preserves anatomical continuity while enabling comprehensive multi-orientation feature integration.
\begin{equation}
\begin{aligned}
QSMamba(z) = & Mamba(z_f) + Mamba(z_r) + \\
             & Mamba(z_i) + Mamba(z_{ri})
\end{aligned}
\end{equation}
where $z_{x}$ denotes scanning strategy.

\subsection{Multi-scale Fusion Mamba Decoder}
\begin{figure}[t]
     \centering
     \includegraphics[width=1\columnwidth]{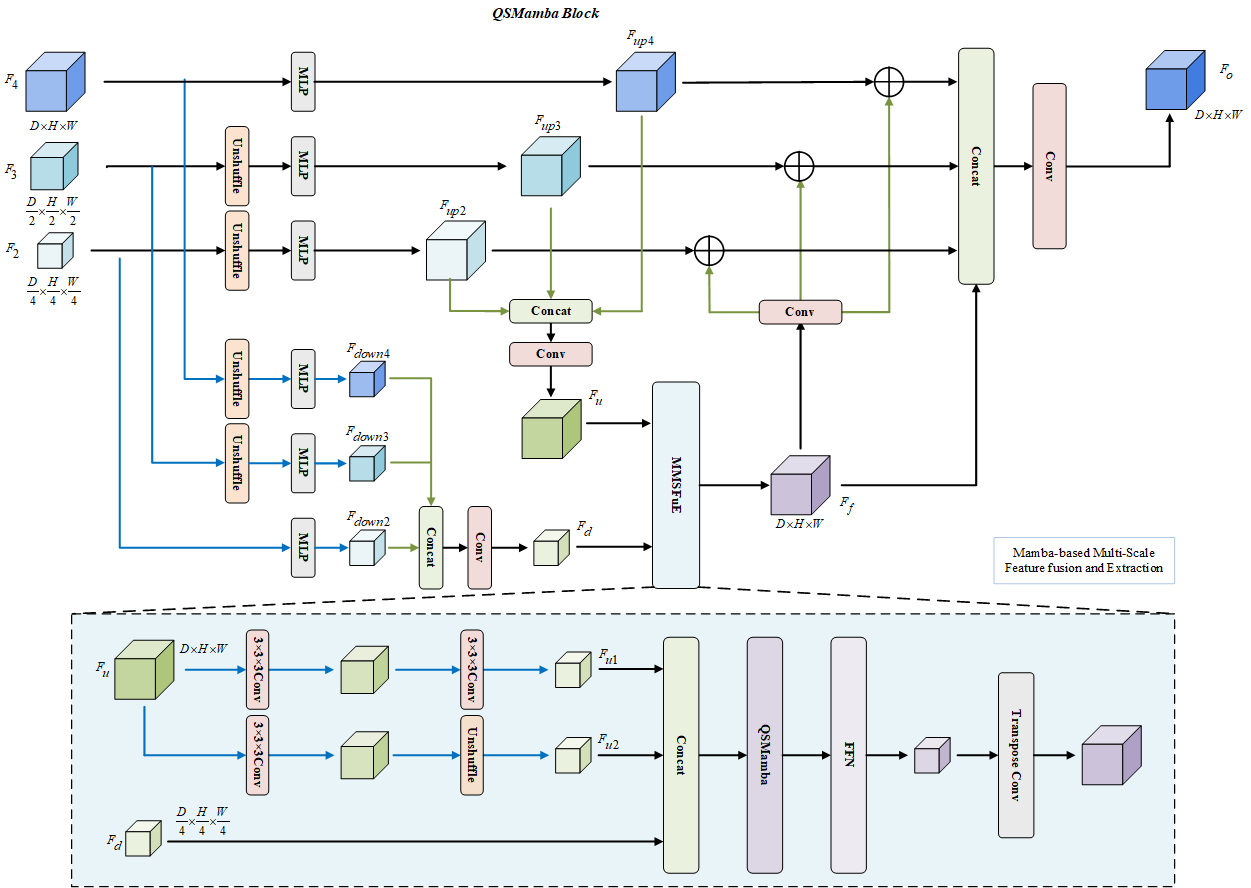}
     \caption{The Mamba decoder architecture with Mamba-based Multi-Scale Feature Fusion and Extraction.}\label{fig3}
\end{figure}

\paragraph{Overview}
As shown in the Figure \ref{fig3}, we directly aggregate the features of different abstraction levels provided by the Mamba encoder, and process them to reconstruct feature maps with sufficient spatial context information. Specifically, the Mamba decoder receives a total of 3 feature maps of different scales, including the resolution \(\frac{1}{2}\) and \(\frac{1}{4}\) feature maps directly output by the Mamba encoder, namely $F_2$, $F_3$, and the feature map $F_4$ after upsampling aggregation. First, $F_2$, $F_3$ and $F_4$ are obtained using linear projection and pixel resampling to get $F_{up2}$, $F_{up3}$, $F_{up4}$ and $F_{down2}$, $F_{down3}$, $F_{down4}$ at the resolution of $F_2$ and $F_4$, respectively. They are then concatenated and processed through a convolutional module with normalization and activation functions to obtain $F_u$ and $F_d$, which are input into our proposed MMSFuE (Mamba-based Multi-Scale Feature Fusion and Extraction) module for processing to extract rich multi-scale context features, thereby generating a new feature map $F_f$. Then, we add the feature map $F_f$ to $F_2$, $F_3$ and $F_4$ to enrich their multi-scale context. The final prediction path connects $F_f$ with these context-enhanced features and outputs the combined features through the convolution module.

\paragraph{Mamba-based Multi-Scale Feature fusion and Extraction}
We proposed a multi-scale context extraction module based on Mamba, whose structure is shown in Figure \ref{fig3}. To effectively reduce the computational cost, we adopted specific strategies for $F_u$: on one hand, we applied a 3D convolution with a stride of 2 twice to downsample $F_u$ by 4 times, obtaining $F_{u1}$; on the other hand, we first downsampled $F_u$ by a stride of 2 using a 3D convolution, and then obtained a 1/4 resolution feature map $F_{u2}$ through pixel resampling technology. The consecutive two $3\times 3\times 3$convolutions not only reduce the computational consumption but also achieve the effect similar to a $5\times 5\times 5$ convolution, thereby obtaining a wider receptive field $F$. The intermediate pixel resampling operation can reduce the scale of the feature map while obtaining the spatial information under a $3\times 3\times 3$ receptive field without loss, further enhancing the extraction effect of context information.

Subsequently, we concatenated $F_d$, $F_{u1}$, and $F_{u2}$ along the channel dimension and input the concatenated feature map into the QSMamba module. This processing method shows significant advantages in speed and efficiency compared to the method of separately performing QSMamba context extraction on $F_u$ and $F_d$. After the above processing, the feature map has a quarter resolution, and to achieve cross-scale context mixing, we performed 4 times upsampling with transposed convolution on it to restore the resolution, ultimately obtaining a new feature map $F_f$ that contains mixed multi-scale context information. Then, we added $F_f$ after convolution processing to $F_4$, $F_3$ and $F_2$, which can further enhance the stage-specific representations in the encoder. From another perspective, we can also view this process as a residual connection of $F_f$ to $F_4$, $F_3$ and $F_2$. This design can effectively inject multi-scale context information into the stage-specific feature maps, enabling the feature maps to inherit rich information from different abstraction levels.
Finally, in the decoding stage, we directly connected the different stage-enhanced context feature maps provided by the Mamba decoder with the multi-scale context feature $F_f$. For different cascaded features, we output through convolution fusion for further processing, providing higher-quality and more discriminative feature representations for subsequent tasks.

\section{Experiments}
\subsection{Dataset}
\paragraph{Synapse}
Synapse Multi-organ segmentation dataset \cite{landman2015miccai}, which was published within the MICCAI 2015 Multi-Atlas Abdomen Labeling Challenge. This dataset includes 3779 axial contrast-enhanced abdominal CT images from 30 abdominal CT scans, with each volume consisting of 85 to 198 slices. It includes a total of thirteen organs in the abdomen: the spleen, the right kidney, the left kidney, the gallbladder, the esophagus, the liver, the stomach, the aorta, the inferior vena cava, the portal vein and the splenic vein, the pancreas, the right adrenal gland, and the left adrenal gland.
\paragraph{BraTS2023}
The Brain Tumor Segmentation Challenge is jointly organized by the Medical Image Computing and Computer Assisted Intervention Society (MICCAI) \cite{kazerooni_BrainTumorSegmentation_2024,menze_MultimodalBrainTumor_2015}. The dataset contains 1251 sets of 3D brain MRI scans. Each case integrates four complementary pulse sequences (T1, T1Gd, T2, T2-FLAIR) with pixel-level annotations delineating three distinct glioma subregions: Whole Tumor (WT), Enhancing Tumor (ET), and Tumor Core (TC).

\begin{table}[t]
     \centering
     \small
     \caption{Segmentation accuracy of different methods on the Synapse dataset.Bold indicates the best option, and underline indicates the second-best option.}
     \begin{tabular}{l|cccccccccccc|cc}
         \hline
         Methods & Spl & Rkid & Lkid & Gall & Eso & Liv & Sto & Aor & IVC & Veins & Pan & AG & Dice & HD \\
         \hline
         SegResNet \cite{myronenko_3DMRIBrain_2018} & 95.26 & 93.94 & 93.77 & 68.76 & 70.65 & 96.10 & 86.06 & 88.94 & 82.80 & 70.29 & 79.29 & 0.00 & 71.11 & 43.74 \\
         SegFormer3D \cite{perera_SegFormer3DEfficientTransformer_2024} & 94.13 & 92.47 & 92.72 & 68.01 & 70.07 & 95.74 & 83.52 & 86.91 & 81.88 & 64.54 & 72.95 & 57.43 & 78.29 & 12.00 \\
         UNETR \cite{hatamizadeh_UNETRTransformers3D_2021} & 94.67 & 92.18 & 92.59 & \textbf{84.41} & 74.33 & 93.10 & \textbf{90.69} & 90.22 & 82.35 & 74.81 & \underline{82.67} & 69.97 & 84.00 & 5.81 \\
         SwinUNETR \cite{cao_SwinUnetUnetlikePure_2021} & \textbf{96.01} & \underline{94.45} & \underline{94.40} & 83.06 & \underline{77.25} & \textbf{96.87} & 87.32 & \textbf{90.57} & \underline{84.13} & \underline{75.29} & 81.30 & 71.21 & 84.85 & 5.16 \\
         SegMamba \cite{xing_SegMambaLongrangeSequential_2024} & 95.43 & 94.08 & 93.98 & 83.49 & 75.61 & \textbf{96.87} & 89.58 & \underline{90.55} & \textbf{86.45} & 73.63 & 81.53 & \textbf{72.51} & \underline{85.09} & \underline{4.01} \\
         \hline
         Ours & \underline{95.98} & \textbf{94.69} & \textbf{94.57} & \underline{83.68} & \textbf{77.83} & \underline{96.75} & \underline{90.24} & 89.70 & 83.75 & \textbf{76.36} & \textbf{83.62} & \underline{71.79} & \textbf{85.44} & \textbf{3.65} \\
         \hline
     \end{tabular}
     \label{tab1}
 \end{table}

 \begin{figure}[t]
     \centering
     \includegraphics[width=1\columnwidth]{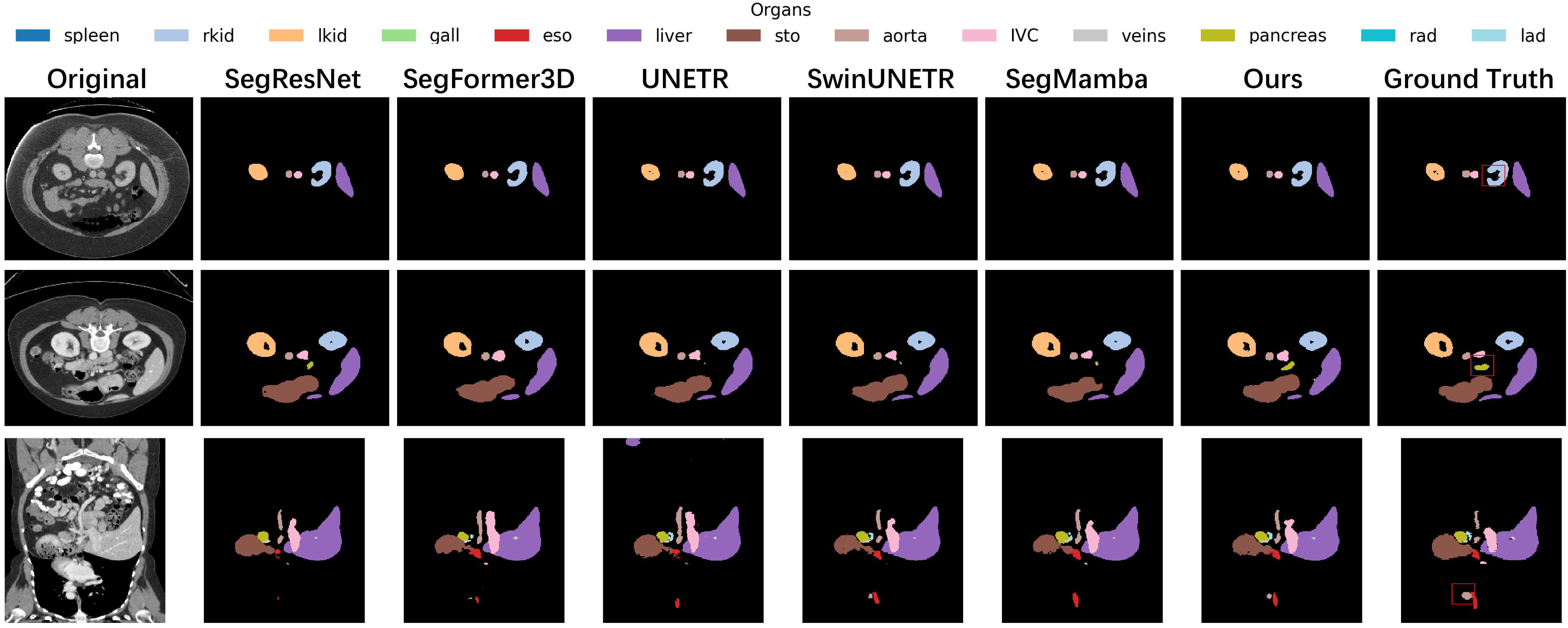}
     \caption{Visual comparison of the segmentation of Synapse multi-organ CT. The red boxes highlight the key details.}\label{v1}
\end{figure}

\subsection{Implementation details}
Our model was implemented using PyTorch 2.3.0-cuda11.8 and Monai 1.3.0 on two NVIDIA GeForce RTX 4090 GPUs. During the training process, for each GPU of the Synapse dataset, we used random cropping with a size of $96 \times 96 \times 96$, and trained for 2000 epochs. For each GPU of the BraTS2023 dataset, we used random cropping with a size of $128 \times 128 \times 128$, and trained for 1000 epochs. The batch size for all datasets was set to 1. In all experiments, we used cross-entropy loss and employed the SGD optimizer with a polynomial learning rate scheduler (initial learning rate of 1e-4, decay rate of 1e-5). For each dataset, we randomly allocated 70\% of the 3D data for training, 10\% for validation, and the remaining 20\% for testing.

\subsection{Compared Methods}
We compare DM-SegNet with six state-of-the-art segmentation methods: a CNN-based approach (SegresNet \cite{myronenko_3DMRIBrain_2018}), three Transformer variants (Segformer3D \cite{perera_SegFormer3DEfficientTransformer_2024}, UNETR \cite{hatamizadeh_UNETRTransformers3D_2021}, SwinUNETR \cite{cao_SwinUnetUnetlikePure_2021}), and a Mamba-driven architecture (SegMamba \cite{xing_SegMambaLongrangeSequential_2024}). To ensure methodological fairness and reproducibility, all comparative models were retrained using author-official implementations under identical training protocols, including standardized data splits, preprocessing pipelines, and hyperparameter configurations. The Dice Similarity Coefficient (DSC) to assess volumetric overlap and the 95th percentile Hausdorff Distance (HD95) to quantify boundary delineation precision, thereby adopted for quantitative comparison on the Synapse and BraTS2023 datasets.
\paragraph{Experiment results on Synapse dataset}
As shown in Table 1, the proposed DM-SegNet demonstrates superior performance compared to previous state-of-the-art methods on the Synapse multi-organ CT dataset. Experimental results indicate that the Dual-Mamba architecture achieves a DSC of 85.44\% and a HD95 of 3.65 mm. The model attains the highest DSC rankings for the Kidney, Esophagus, Portal Vein and Splenic Vein, and Pancreas, while achieving second-best DSC performance for the Spleen, Gallbladder, Liver, Stomach, and Adrenal Gland. This consistent performance hierarchy across both solid and hollow organ systems underscores the architecture's robustness in addressing diverse anatomical characteristics and intensity variations inherent in abdominal CT imaging.
\paragraph{Experiment results on BraTS2023 dataset}
Our DM-SegNet achieved the highest DSC of 90.53\%, 93.13\%, and 86.99\% in TC, WT, and ET, respectively, with a mean DSC of 90.22\%.  It also obtained the lowest HD95 of 3.64 mm, 3.81 mm, and 3.44 mm in TC, WT, and ET, respectively, and a mean HD95 of 3.63 mm, demonstrating superior generalization ability and segmentation robustness.

\begin{table}[t]
    \centering
    \caption{Comparison of different methods. Bold indicates the best option, and underline indicates the second-best option.}
    \begin{tabular}{l|cc|cc|cc|cc}
        \hline
        \multirow{2}{*}{Methods} & \multicolumn{2}{c}{TC} & \multicolumn{2}{c}{WT} & \multicolumn{2}{c}{ET} & \multicolumn{2}{c}{Mean} \\
        & Dice & HD & Dice & HD & Dice & HD & Dice & HD \\
        \hline
        SegResNet \cite{myronenko_3DMRIBrain_2018} & 88.63 & 4.22 & 92.21 & 4.30 & 83.86 & 4.55 & 88.24 & 4.36 \\
        SegFormer3D \cite{perera_SegFormer3DEfficientTransformer_2024} & 88.33 & 4.14 & 91.98 & 4.12 & 80.42 & 4.68 & 86.91 & 4.31 \\
        UNETR \cite{hatamizadeh_UNETRTransformers3D_2021} & 86.54 & 4.52 & 91.80 & 4.95 & 82.86 & 4.82 & 87.06 & 4.76 \\
        SwinUNETR \cite{cao_SwinUnetUnetlikePure_2021} & 88.69 & 4.08 & 92.49 & 4.28 & 84.64 & 4.05 & 88.61 & 4.13 \\
        SegMamba \cite{xing_SegMambaLongrangeSequential_2024} & \underline{89.85} & \underline{3.76} & \underline{92.85} & \underline{4.34} & \underline{85.27} & \underline{3.80} & \underline{89.32} & \underline{3.97} \\
        \hline
        Ours & \textbf{90.53} & \textbf{3.64} & \textbf{93.13} & \textbf{3.81} & \textbf{86.99} & \textbf{3.44} & \textbf{90.22} & \textbf{3.63} \\
        \hline
    \end{tabular}
    \label{tab2}
    
    \vspace{1em} % 添加垂直间距
    
    \centering
    \caption{Segmentation performance of ablation study on dataset Synapse. QSS represents the Quadri-directional Scanning Strategy.}
    \begin{tabular}{l|cc|cc}
        \hline
        Methods & GSC & QSS & Dice & HD \\
        \hline
        M1 &  &  & 83.71 & 6.49 \\
        M2 & \checkmark &  & 84.37 & 4.94 \\
        M3 &  & \checkmark & 84.65 & 4.68 \\
        \hline
        Ours & \checkmark & \checkmark & \textbf{85.44} & \textbf{3.65} \\
        \hline
    \end{tabular}
    \label{tab3}
\end{table}

 \begin{figure}[t]
     \centering
     \includegraphics[width=1\columnwidth]{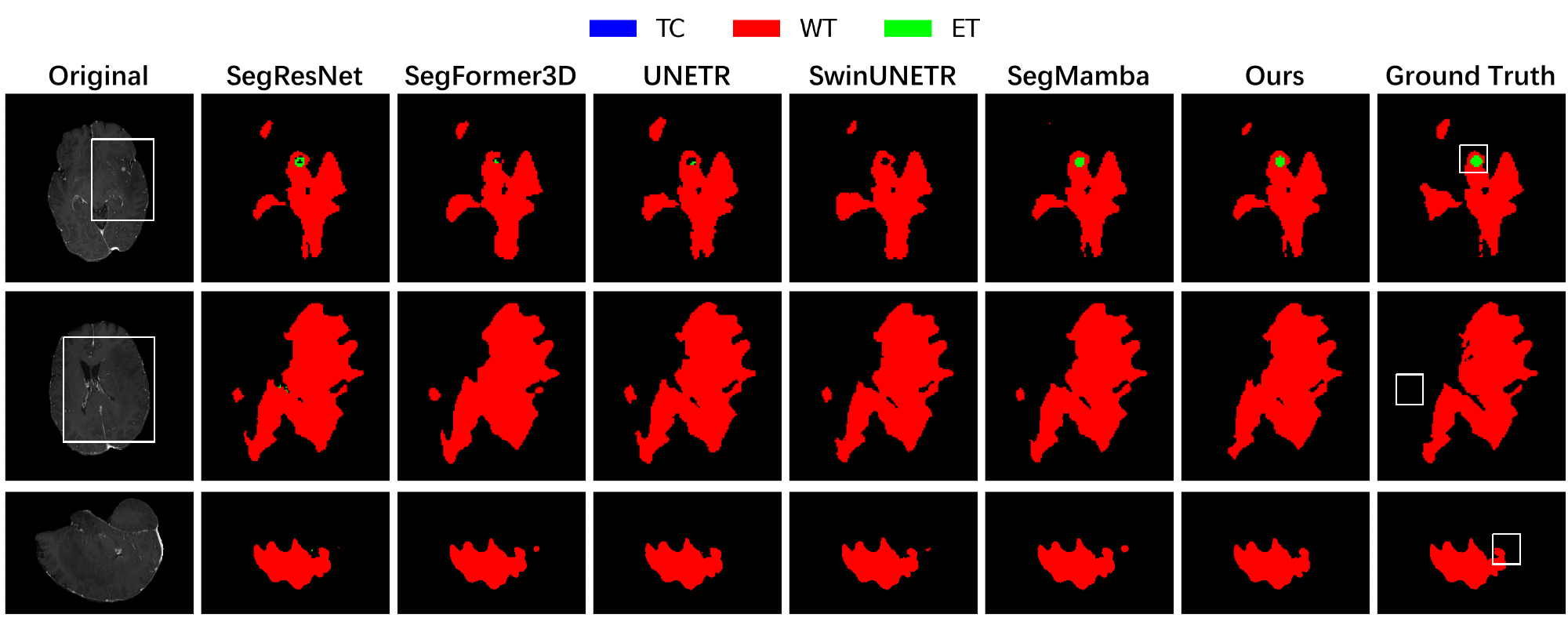}
     \caption{Visual comparison of the segmentation of brain tumor segmentation. White boxes indicate the focus position and highlight critical details.}\label{v2}
\end{figure}

\paragraph{Visual comparisons}
As illustrated in Figure \ref{v1}, our DM-SegNet demonstrates precise segmentation of multiple abdominal organs on the Synapse dataset, exhibiting superior performance in organ boundary delineation compared with other methods. As shown in Figure \ref{v2}, the proposed method also achieves accurate detection of tumor region boundaries on the BraTS2023 dataset. In contrast to existing approaches, our framework not only precisely localizes subtle tumor boundaries but also significantly reduces false positives in normal regions. These segmentation results exhibit better consistency with ground truth annotations compared to state-of-the-art methods.

\subsection{Ablation Study}
As shown in Table 3, we conduct a comprehensive ablation study on the Synapse dataset to evaluate the contribution of key components in our 3D medical image segmentation framework. The quantitative analysis reveals that the baseline (M1) achieves a DSC of 83.71\% with HD95 of 6.49 mm. Method M2, with GSC enabled, improves to 84.37\% DSC and 4.94 mm HD95, demonstrating GSC's positive contribution. The exclusive integration of the Quadri-directional Scanning Strategy (QSS) alone further elevates the DSC to 84.65\% while reducing the HD95 to 4.68 mm, demonstrating that spatial contextual modeling through multi-directional scanning effectively captures fine anatomical structures. Our full framework achieves state-of-the-art performance with 85.44\% DSC and 3.65 mm HD95, outperforming all ablation variants. The 1.73\% absolute DSC gain over the M1 baseline and 43.7\% HD95 reduction quantitatively validate the effectiveness of the proposed modules.

\section{Conclusion}
In this work, we propose DM-SegNet, a dual-Mamba framework for 3D medical image segmentation. By integrating reinforced spatial encoding with global context-aware decoding enhancement, our architecture resolves the inherent encoder-decoder incompatibility in conventional SSM-based approaches. The quadri-directional spatial Mamba module preserves 3D anatomical continuity through multi-directional scanning, eliminating spatial distortion caused by 1D sequence flattening. Coupled with the spatial feature enhancement of gated spatial convolution  layers and a Mamba-driven decoder enabling bidirectional state synchronization, DM-SegNet achieves excellent performance in clinically critical tasks.

Extensive validation on the Synapse and BraTS2023 datasets demonstrates the superiority of our framework, attaining 85.44\% DSC for abdominal organ segmentation and 90.22\% DSC for brain tumor delineation. Our ablation studies quantitatively confirm the synergistic contributions of the QSM module's multi-directional scanning strategy, GSC layers, and bidirectional decoding strategy in enhancing 3D segmentation accuracy and reducing spatial edge errors.

\paragraph{Limitations and Future Work}
While Mamba has evolved significantly over time, its design is inherently tailored for 1D data, necessitating specific adaptations for 3D medical imaging. Although our multi-directional scanning strategy and global context-aware decoding partially alleviate dimensional discrepancies, they remain constrained by computational time and memory footprint, ultimately limiting comprehensive processing of spatial information. Further research is required to address these challenges.
%% The Appendices part is started with the command \appendix;
%% appendix sections are then done as normal sections
%\appendix
%\section{Example Appendix Section}
%\label{app1}
%Appendix text.

%% 使用BibTeX文件
\bibliographystyle{elsarticle-num} 
\bibliography{references}

\end{document}